 \documentclass{aa}
\usepackage{graphicx}
\usepackage{natbib}
\usepackage{txfonts}

\begin{document}

     \title{Chemical sensitivity to the ratio of the cosmic-ray ionization rates of He and H$_2$ in dense clouds}

     \subtitle{ }

     \author{V. Wakelam\inst{1,2} \and E. Herbst\inst{1,3} \and F. Selsis\inst{4} \and G. Massacrier\inst{4}}
     \offprints{V. Wakelam, \email{wakelam@mps.ohio-state.edu}}
     \institute{
      Department of Physics, The Ohio State University, Columbus, OH 43210, USA \and
      Observatoire Aquitain des Sciences de l'Univers, Laboratoire d'Astrodynamique, d'Astrophysique et d'A\'eronomie de Bordeaux, CNRS/INSU UMR 5804, BP 89, 33270 Floirac, France \and 
      Departments of Astronomy and Chemistry, The Ohio State University, Columbus, OH 43210, 		USA \and Ecole Normale Sup\'erieure de Lyon, Centre de Recherche Astronomique de Lyon,
46 all\'ee d'Italie, 69364 Lyon Cedex 07, France ; CNRS, UMR 5574 ;
Universit\'e de Lyon 1, Lyon, France.   }

     \date{Received 21 April 2006  / Accepted 9 August 2006 }

      \abstract
{}
 {To determine whether or not gas-phase chemical models with homogeneous and time-independent physical conditions explain the many observed molecular abundances in astrophysical sources, it is crucial to estimate the uncertainties in the calculated abundances and compare them with the observed abundances and their uncertainties. Non linear amplification of the error and bifurcation may limit the applicability of chemical models. Here we study such effects on dense cloud chemistry. }
 { Using a previously studied approach to uncertainties based on the representation of rate coefficient errors as log normal distributions, we attempted to apply our approach using as input a variety of different elemental abundances from those studied previously. In this approach, all rate coefficients are varied randomly within their log normal (Gaussian) distribution, and the time-dependent chemistry calculated anew many times so as to obtain good statistics for the uncertainties in the calculated abundances.}
 {Starting with so-called ``high-metal'' elemental abundances, we found bimodal rather than Gaussian like distributions for the abundances of many species and traced these strange distributions to an extreme sensitivity of the system to changes in the ratio of the cosmic ray ionization rate $\zeta_{\rm He}$ for He and that for molecular hydrogen $\zeta_{\rm H_2}$.  The sensitivity can be so extreme as to cause a region of bistability, which was subsequently found to be more extensive for another choice of elemental abundances.  To the best of our knowledge, the bistable solutions found in this way are the same as found previously by other authors, but it is best to think of the ratio $\zeta_{He}/\zeta_{H_{2}}$ as a control parameter perpendicular to the ''standard'' control parameter $\zeta/n_{\rm H}$. } 
 {}


     \titlerunning{Extreme sensitivity of the dense cloud chemistry}

     \maketitle
    
\section{Introduction}

The equations describing the chemical evolution of quiescent cores (also known as molecular clouds) are highly 
non-linear and can result in extreme sensitivity to the initial conditions or to some of the parameters. For example, the elemental ratio of carbon to oxygen is well known to be an important parameter for dense cloud chemistry \citep{1998ApJ...501..207T}. Another well-known consequence of this non-linearity is the presence of bistable regions for  some ranges of model parameters (temperature, density, comic-ray ionization rate, rate coefficients and elemental abundances).  The phenomenon of bistability  is characterized by two stable solutions for chemical abundances at steady-state for the same set of parameters. Bistability in dark cloud conditions, i.e. at low temperature, was first discovered by \citet{1993ApJ...416L..87L} and subsequently studied by a large number of authors \citep{1995A&A...297..251L,1995A&A...296..779S,1998A&A...334.1047L,2000RSPTA.358.2549P,2003A&A...399..583C,2006ApJ...645..314B}. The two solutions of this bistability are characterized by a large difference in the C/CO and O/O$_2$ ratios and in the ionization fraction. For this last reason, they were named the High and Low Ionization Phases (HIP and LIP). \citet{1995A&A...297..251L} showed that the region of bistability can be mapped out by variations in density, temperature, cosmic-ray ionization rate, and elemental depletions within specific ranges.   \citet{2000RSPTA.358.2549P} later showed that even variations in rate coefficients can lead to the two solutions. Recently, \citet{2006ApJ...645..314B} identified the chemical mechanisms of this phenomenon as a cycle involving the H$^+_3$, O$_2$, and S$^+$ species. 

In two previous papers, we presented a Monte Carlo method to compute the theoretical error of abundances in chemical models due to uncertainties in rate coefficients and physical conditions \citep{2005A&A...444..883W,2006A&A...451..551W}. The errors in the rate coefficients are defined by a log-normal distribution. As a consequence, the resulting abundances  at steady-state usually follow a Gaussian distribution \citep{2005A&A...444..883W}. When applying this method, we found that the chemical abundances are very sensitive to the ratio between the ionization rates of helium $\zeta_{\rm He}$ and molecular hydrogen $\zeta_{\rm H_2}$ caused by cosmic-ray particles. Characterized by a variation of several orders of magnitude in some abundances for small variations of $\zeta_{\rm He} / \zeta_{\rm H_2}$, this sensitivity can lead to bimodal rather than Gaussian distributions for the abundances. For certain ranges of the parameters, it can even lead to bistability, with the ratio $\zeta_{\rm He} / \zeta_{\rm H_2}$ appearing to be a control parameter. 

This paper, which describes the sensitivity and some of its consequences, is organized as follows. In section 2, we briefly present the chemical model and the uncertainty method used for this analysis. In a somewhat unorthodox way, we have decided to present, in section 3, the manner in which the sensitivity was found while studying the use of so-called ``high-metal'' elemental abundances. Section 4 shows the influence of the elemental abundances and the chemical network used  on the sensitivity. In section 5, we present and discuss the phenomenon of bistability as obtained with the variation of $\zeta_{\rm He} / \zeta_{\rm H_2}$. In section 6,   the chemical reactions involved in this sensitivity/bistability are elucidated. Finally, the last section presents our conclusions.

\section{Chemical model and method of uncertainty}\label{model}


\begin{table}
\caption{High- and low- metal elemental abundances with respect to H$_2$.  }
\begin{center}
\begin{tabular}{lcc}
\hline
\hline
Element & High-Metal & Low-Metal \\
 & & (Intermediate) \\
\hline
He & $2.8\times 10^{-1}$ & $2.8\times 10^{-1}$ \\
N & $4.28\times 10^{-5}$ & $4.28\times 10^{-5}$ \\
O & $3.52\times 10^{-4}$ & $3.52\times 10^{-4}$  \\
C$^+$ & $1.46\times 10^{-4}$ & $1.46\times 10^{-4}$  \\
S$^+$ & $1.6\times 10^{-5}$ &  $1.6\times 10^{-7}$ \\
 & & ($2.0\times 10^{-6}$) \\
Si$^+$ & $1.6\times 10^{-6}$ &  $1.6\times 10^{-8}$ \\
Fe$^+$ & $6.0\times 10^{-7}$ & $6.0\times 10^{-9}$ \\
Na$^+$ & $4.0\times 10^{-7}$ & $4.0\times 10^{-9}$ \\
Mg$^+$ & $1.4\times 10^{-7}$ & $1.4\times 10^{-8}$ \\
P$^+$ & $6.0\times 10^{-7}$ & $6.0\times 10^{-9}$ \\
Cl$^+$ & $8.0\times 10^{-7}$  & $8.0\times 10^{-9}$ \\
\hline
\end{tabular}
\begin{list}{}{}
\item Note: The intermediate-metal case differs from the low-metal case  by the S$^+$ abundance only, as is indicated in brackets in the table.
\end{list}
\end{center}
\label{elem_abund}
\end{table}

We used a gas-phase time-dependent chemical model with the  osu.2003\footnote{http://www.physics.ohio-state.edu/$\sim$eric/research\_files/} network reported by \citet{2004MNRAS.350..323S} (4233 reactions, 421 species and 12 elements). The model computes the evolution of the species for a fixed temperature and density. The initial conditions are all atomic except for molecular hydrogen, while the chemical model and database are the same as used in \citet{2006A&A...451..551W}. For this work, we will present the results using three different sets of elemental abundances: the high-, low- and intermediate-metal cases. The high- and low-metal elemental abundances have been defined by \citet{1982ApJS...48..321G} and are listed in Table~\ref{elem_abund}. The intermediate case has the same abundances as the low-metal one except that the amount of the element sulfur is raised to $2\times 10^{-6}$ compared with H$_2$  (also given in Table~\ref{elem_abund}).  
The other parameters are the typical ones for quiescent cores: a kinetic temperature of 10~K, an H$_2$ density of $10^4$~cm$^{-3}$,  a visual extinction of 10 so that the photochemistry driven by the external UV photons does not occur,  and a fixed cosmic-ray ionization rate $\zeta$ of $1.3\times 10^{-17}$~s$^{-1}$. 

The method used in this analysis to derive uncertainties in abundance is described in detail in \citet{2005A&A...444..883W}. The uncertainties in rate coefficients derive from the UMIST database when available, but for most of the reactions, including the ionization rates of H$_2$ and He, we have assumed a factor of 2 uncertainty. Note that the uncertainty in ionization rates is not an uncertainty in the parameter $\zeta$, but an uncertainty in the multiplicative factor that distinguishes the individual rates.  Although ionization rates for H$_{2}$ and He have been used in a large number of models, the current rates used in terms of $\zeta$ derive back to the Ph. D. thesis of \citet{1975PhDT........71B}, where the factors are 0.93 for the production of H$_{2}^{+}$ ($\zeta_{\rm H_2}^0$) and 0.5 for the production of He$^{+}$ ($\zeta_{\rm He}^0$), leading to a ratio $\zeta_{\rm He}^0 / \zeta_{\rm H_2}^0$ of 0.54, if one neglects the minor channels for H$_{2}$. An older ratio of unity was used by \citet{1973ApJ...185..505H}.   New estimates, involving both direct ionization and secondary ionization by energetic electrons, are sorely needed (Black, private communication) based on the treatment of \citet{1999ApJS..125..237D} for  a mixture of atomic and molecular hydrogen and helium.

The uncertainty method consists of generating $N$ new sets of rate coefficients by replacing each coefficient $k_i$ by a random value consistent with its uncertainty factor $F_i$.   We assume a normal distribution of $\log k_i$ with a standard deviation $\sigma_i = \log F_i$.  We run the model for each set $j$, which produces, for each species, $N$  values of the fractional abundances $X_j (t)$ at a time $t$. For this work, we ignore the uncertainty in physical conditions and consider the uncertainty in the rate coefficients only.  A total $N$ of 2000 different runs was made for each set of parameters studied; this number is large enough to achieve  statistical significance \citep[see][]{2006A&A...451..551W}.

\section{The sensitivity to $\zeta_{\rm He} / \zeta_{\rm H_2}$ in the high-metal case}



\begin{figure}
\begin{center}
\includegraphics[width=1\linewidth]{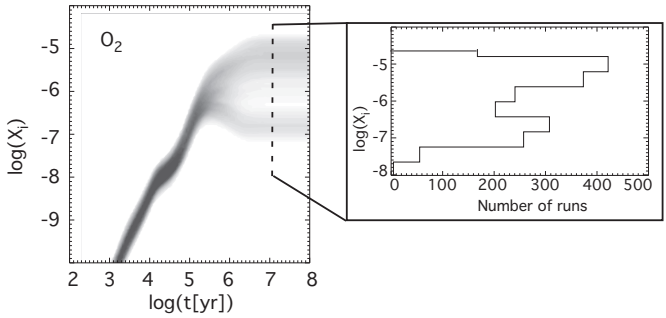}
\caption{Density of probability of the O$_2$ abundance as a function of time (left panel). The right panel shows the histogram of the abundance at $10^7$~yr. The elemental abundances are for the high-metal case. }
\label{density_dist_O2}
\end{center}
\end{figure}

\begin{figure}
\begin{center}
\includegraphics[width=1\linewidth]{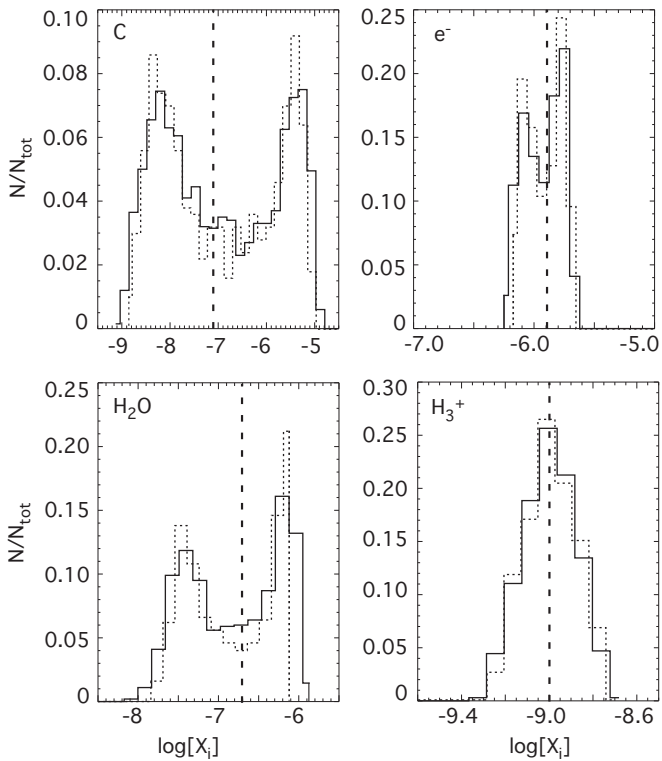}
\caption{Distributions (nomalized to the total number of runs) of the abundances of C, H$_2$O, e$^-$ and H$_3^+$ at $10^7$~yr (high-metal case). The solid lines refer to the variation of all the rate coefficients whereas the dotted lines represent the variation of $\zeta_{\rm He}$ and $\zeta_{\rm H_2}$ only. The vertical dashed line represents the abundance computed with the standard values of the rate coefficients.}
\label{4dist}
\end{center}
\end{figure}

The computation of the theoretical error due to the uncertainties in rate coefficients was first applied to the case of high-metal elemental abundances. 
Figure~\ref{density_dist_O2} shows the results of the random variation of all the rate coefficients for the O$_2$ abundance as a function of time. As can be seen from this figure, the distribution of the abundance is spread into two peaks after $10^6$ yr, as steady-state approaches,  with a significant number of curves in each peak and some stable solutions between the two peaks. The bimodality of the distribution at a specific time (10$^{7}$ yr) is also shown as a histogram in Figure~\ref{density_dist_O2}. Such bimodal distributions are obtained for a large number of species (C, H$_2$O, SO, CS etc) with the largest separation between the two peaks for atomic carbon (three orders of magnitude). Fig.~\ref{4dist} shows histograms for the abundance distributions of the species C, H$_{2}$O, e$^{-}$, and H$_{3}^{+}$ at 10$^{7}$ yr. It can be seen that the H$_3^+$  distribution does not show a bimodal profile but is Gaussian.  The same is true for H$_2$S and OH. Since the error in the rate coefficients follows a log-normal distribution, the bimodal shapes are due to non-linear effects. 


\begin{figure}
\begin{center}
\includegraphics[width=1\linewidth]{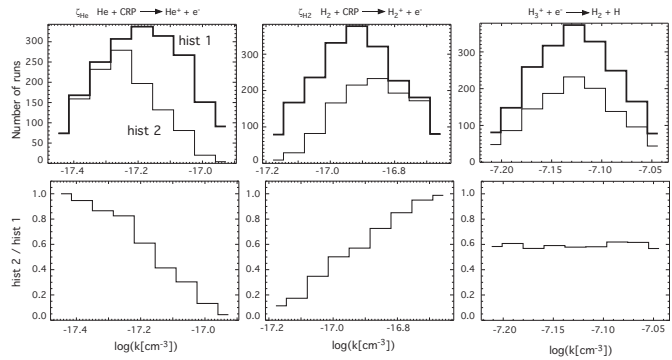}
\caption{Histograms of the rate coefficients of the following reactions (upper panel): He + CRP $\rightarrow$ He$^+$ + e$^-$ ($\zeta_{\rm He}$, left box), H$_2$ + CRP $\rightarrow$ H$_2^+$ + e$^-$ ($\zeta_{\rm H_2}$, middle box) and H$_3^+$ + e$^-$ $\rightarrow$ H$_2$ + H (right box). The thick and thin lines represent the histograms of all the runs (hist 1) and of the runs giving the higher abundance of O$_2$ (hist 2, see text) respectively. The lower panels show the ratio between hist 2 and hist 1. }
\label{dist_k}
\end{center}
\end{figure}

We obtained the bimodal distributions by varying the rate coefficients, we then decided to identify the reactions responsible for them if at all possible. We proceeded as follows.
Among the 2000 runs, we listed the sets of reactions that give one the higher of the two most probable abundances for O$_2$. Then for each rate coefficient, we plotted the histograms for all runs  and the histograms for those runs giving only the higher abundance of O$_2$.  In the upper panels of Fig.~\ref{dist_k}, pairs of histograms can be seen for the rate coefficient of the cosmic ray ionization of He, that of H$_{2}$, and the dissociative recombination of H$_{3}^{+}$ to form H$_{2}$ and H.  The thick lines (hist1) represent histograms for all runs while the thin lines (hist2) represent those ending up with the higher O$_{2}$ abundance.  For reactions not responsible for the bimodal distributions,  there should be no difference between the pairs of histograms except that the total number of runs should be lower for the second one. For most of the reactions, we indeed found no difference, as for the reaction H$_3^+$ + e$^-$ $\rightarrow$ H$_2$ + H in Fig.~\ref{dist_k}.  In fact, there were only two exceptions: the direct ionization rate coefficients of He and H$_2$ by  cosmic rays ($\zeta_{\rm He}$ and $\zeta_{\rm H_2}$), and their marked effect can be seen in Fig.~\ref{dist_k}.  
The high abundance of O$_2$ tends to be formed with lower values of $\zeta_{\rm He}$ and higher values of $\zeta_{\rm H_2}$, as can be seen most clearly in the lower panels of Fig.~\ref{dist_k}, in which the ratios of the two histograms are plotted.   It is then convenient to consider the sensitivity of the abundance of O$_2$ and other species to the ratio $\zeta_{\rm He}$/$\zeta_{\rm H_2}$ rather than to their absolute values. 
To confirm their importance, we randomly varied only $\zeta_{\rm He}$ and $\zeta_{\rm H_2}$, keeping the other rate coefficients constant,  and we obtained the same bimodal distribution as with the fully random method at steady state. The dispersion of the abundances in this case is then almost as large as when we varied all the rate coefficients, as can be seen in the dotted lines in Fig.~\ref{4dist}. 


\begin{figure}
\begin{center}
\includegraphics[width=0.7\linewidth]{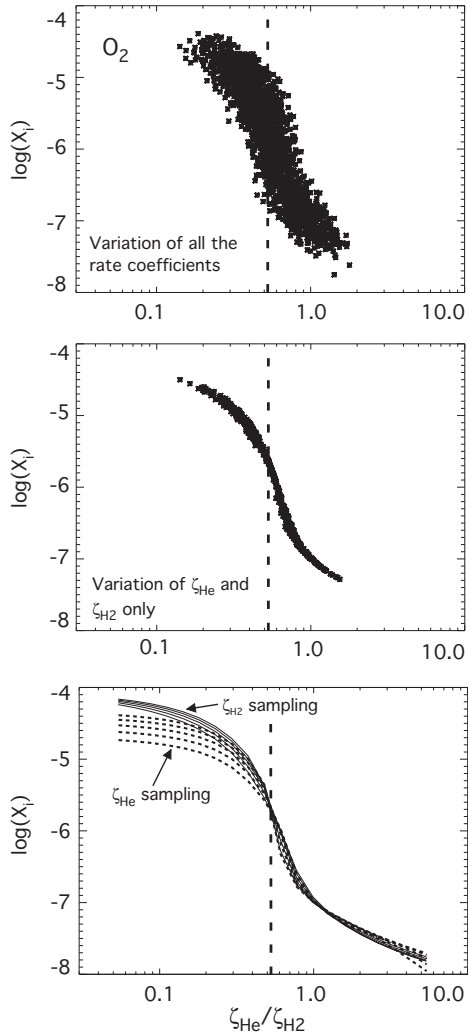}
\caption{Abundance of O$_2$ as a function of $\zeta_{\rm He}$/$\zeta_{\rm H_2}$ at $10^7$~yr. The upper and middle panels represent the abundance computed with the random distribution of all the rate coefficients and of $\zeta_{\rm He}$ and $\zeta_{\rm H_2}$ only, respectively. The solid lines in the lower plot have been computed using one value of $\zeta_{\rm He}$ for each line and varying $\zeta_{\rm H_2}$ , while the dotted lines have been computed  using one value of $\zeta_{\rm H_2}$ for each line and varying  $\zeta_{\rm He}$.  The vertical dashed line indicates the standard value of $\zeta_{\rm He}^0$/$\zeta_{\rm H_2}^0$.}
\label{k1k2_abO2}
\end{center}
\end{figure}

\begin{figure}
\begin{center}
\includegraphics[width=1\linewidth]{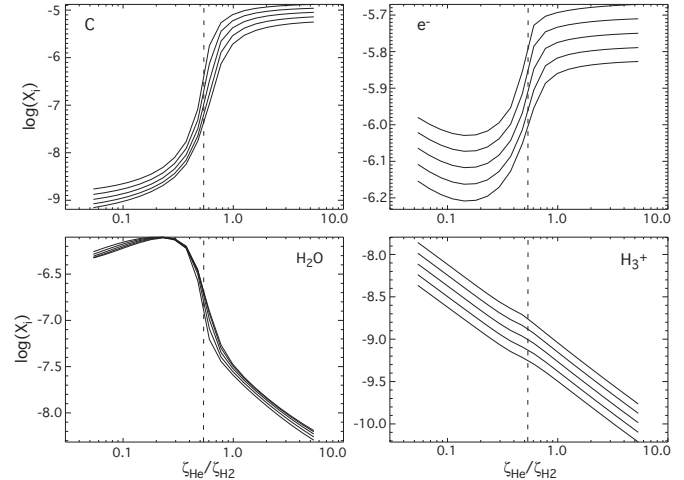}
\caption{Abundances of C, e$^-$, H$_2$O and H$_3^+$ as a function of $\zeta_{\rm He}$/$\zeta_{\rm H_2}$ and for five different values of $\zeta_{\rm He}$ (solid lines). The vertical dashed line indicates the standard value of $\zeta_{\rm He}^0$/$\zeta_{\rm H_2}^0$.}
\label{4k1k2_echant}
\end{center}
\end{figure}

The sensitivity of the chemical abundances to the ratio $\zeta_{\rm He}$/$\zeta_{\rm H_2}$ can be explored in another way. In Fig.~\ref{k1k2_abO2} (left panel), we show the steady-state abundance of O$_2$, computed by randomly varying all the rate coefficients  as a function of $\zeta_{\rm He}$/$\zeta_{\rm H_2}$. For comparison, the abundance computed with the random variation of $\zeta_{\rm He}$ and $\zeta_{\rm H_2}$ only is shown in the middle plot of the figure. The smaller vertical dispersion is caused by not varying the other rate coefficients.  Finally and for more clarity, we depict, in the lower panel, the calculated O$_{2}$ abundance as a function of  $\zeta_{\rm He}$/$\zeta_{\rm H_2}$ obtained by calculations in which the ionization rates are not varied randomly.  Rather, the solid lines show values of the steady-state O$_2$ abundance computed for 5 values of $\zeta_{\rm He}$ (between 2$\zeta_{\rm He}^0$ and $\zeta_{\rm He}^0$/2) with  $\zeta_{\rm H_2}$ varied to account for the range of the abscissa between 0.05 and 5.  In the same panel,  dotted lines show the O$_2$ abundances computed for 5 values of $\zeta_{\rm H_2}$ with  $\zeta_{\rm He}$ varied to account for the range of the abscissa. The abundances computed by each method  are similar especially at the inflection point, confiming that the choice of the ionization ratio as a parameter of the sensitivity is a reasonable one.  

The abundance of O$_2$ decreases by two orders of magnitude (from $10^{-5}$ to $10^{-7}$) when $\zeta_{\rm He}$/$\zeta_{\rm H_2}$ goes from 0.3 to 0.8, and the inflection point of the curve occurs at the standard value of $\zeta_{\rm He}^0$/$\zeta_{\rm H_2}^0$ used in osu.2003 (0.5/0.93). 
In Fig.~\ref{4k1k2_echant} we show some other examples of this sensitivity by using five different values of  $\zeta_{\rm He}$, as in the solid lines of the right panel of Fig.~\ref{k1k2_abO2}. The abundance of atomic carbon has an almost equal sensitivity albeit opposite to that of O$_2$, while the H$_2$O abundance shows a similar profile to O$_2$ except for a local maximum at $\zeta_{\rm He}$/$\zeta_{\rm H_2}$ around 0.2.   Although the electronic abundance increases towards higher $\zeta_{\rm He}$/$\zeta_{\rm H_2}$, it varies only a factor of 2.5 between $\zeta_{\rm He}$/$\zeta_{\rm H_2}$=0.1 and $\zeta_{\rm He}$/$\zeta_{\rm H_2}$=3. The abundance of the H$_3^+$ ion, which does not show any bimodal distribution, is linear with $\zeta_{\rm He}$/$\zeta_{\rm H_2}$. The dispersion of the abundances, for a given ratio $\zeta_{\rm He}$/$\zeta_{\rm H_2}$, due to the variation of $\zeta_{\rm He}$ by a factor of 2 around the standard value $\zeta_{\rm He}^0$, depends on the species: it is smaller for H$_2$O, O$_2$ and e$^-$ than for C and H$_3^+$, for instance, but remains anyway below 0.5 dex. 

\section{Sensitivity for different model parameters}
\subsection{Other choices of elemental abundances}

\begin{figure}
\begin{center}
\includegraphics[width=1\linewidth]{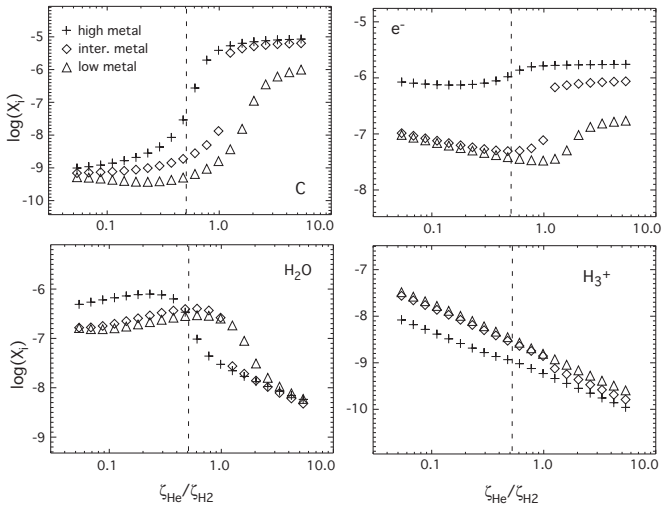}
\caption{Steady-state abundances of C, e$^-$, H$_2$O and H$_3^+$ as  functions of $\zeta_{\rm He}$/$\zeta_{\rm H_2}$ for three different elemental abundances.  The triangles, diamonds and crosses represent the low-, intermediate-, and high-metal cases. The vertical dashed line indicates the standard value of $\zeta_{\rm He}^0$/$\zeta_{\rm H_2}^0$. The H$_2$ density is $10^4$~cm$^{-3}$ and the temperature 10~K. The value of $\zeta_{\rm He}$ is  fixed at its standard value while  $\zeta_{\rm H_2}$ is varied. }
\label{k1k2_3CI}
\end{center}
\end{figure}

\begin{figure}
\begin{center}
\includegraphics[width=1\linewidth]{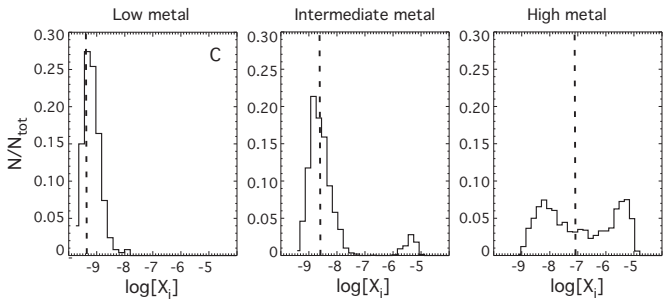}
\caption{Distribution (nomalized with respect to the total number of runs) of the steady-state abundance of atomic carbon for the three sets of elemental abundances. The vertical dashed line represents the abundance computed with the standard values of the rate coefficients. }
\label{dist_C_3CI}
\end{center}
\end{figure}

The calculated sensitivity depends on the elemental abundances used for the modeling. In Fig.~\ref{k1k2_3CI}, we show the C, H$_2$O, e$^-$ and H$_3^+$ fractional abundances as a function of $\zeta_{\rm He}$/$\zeta_{\rm H_2}$ for the low-, intermediate-, and high-metal elemental abundances (see Sect.~\ref{model}). The sensitivity plots, as well as most others later in the paper, are obtained by fixing the helium ionization rate at its standard value so that no dispersion is seen. The H$_2$ density remains at $10^4$~cm$^{-3}$ and the temperature at 10~K. Although the sensitivity exists for the three cases, the inflection point is shifted to higher values of $\zeta_{\rm He}$/$\zeta_{\rm H_2}$ for the low-metal case. For the intermediate case, on the other hand, the sensitivity seems to be stronger, leading to an abrupt jump in the abundance of C for $\zeta_{\rm He}$/$\zeta_{\rm H_2}$=1.1. A similar abrupt change can be seen for the electronic abundance.  This jump is actually a manifestation of bistability  (see section~\ref{bistability}). Note that for $\zeta_{\rm He}$/$\zeta_{\rm H_2} < 1.1$, the abundances for the intermediate-metal case tend to be close to the low-metal ones whereas for $\zeta_{\rm He}$/$\zeta_{\rm H_2} > 1.1$ they are similar to the high-metal ones. 

When one runs the uncertainty method with a factor of two uncertainty in $\zeta_{\rm He}$ and $\zeta_{\rm H_2}$, the values of $\zeta_{\rm He}$/$\zeta_{\rm H_2}$ are spread approximatively between 0.13 and 2.2. The distribution of the abundances computed with this method thus depends on the sensitivity within this range of values. The distribution of the steady-state abundances of C computed by variation of all the rate coefficients  are shown in Fig.~\ref{dist_C_3CI} for the three sets of abundances. Because the inflection point for the low-metal case is shifted to $\zeta_{\rm He}$/$\zeta_{\rm H_2}$=2, the uncertainty method gives Gaussian-like distributions although slightly asymmetrical towards higher C abundances. This explains why we did not see the sensitivity effect in our previous work on rate coefficient uncertainties in dark clouds  \citep[see][]{2006A&A...451..551W}. For high-metal elemental abundances, the point of inflection occurs close to the standard value of $\zeta_{\rm He}$/$\zeta_{\rm H_2}$=0.54 and the distribution for the C abundance is bimodal.  When using the intermediate-metal abundances, the uncertainty method gives two distinct peaks with no solutions between them. The peak at higher abundance is the smaller since the point of inflection occurs at values of $\zeta_{\rm He}$/$\zeta_{\rm H_2}$=1.1 somewhat larger than the standard. The lack of any model runs with steady-state abundances in between the two distributions stems from the sharpness of the curve around the inflection point and is a consequence of the bistability.


\begin{figure}
\begin{center}
\includegraphics[width=1\linewidth]{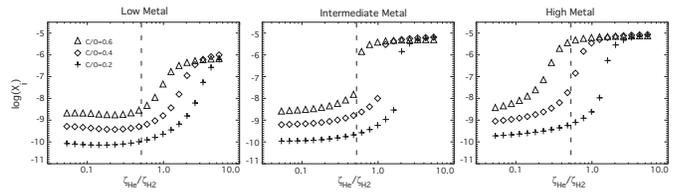}
\caption{Atomic carbon abundance at steady state as a function of $\zeta_{\rm He}$/$\zeta_{\rm H_2}$ for the three elemental abundances (low-, intermediate-, and high-metal). In each plot, the symbols refer to different C/O ratios. The vertical dashed line indicates the standard value of $\zeta_{\rm He}^0$/$\zeta_{\rm H_2}^0$. }
\label{k1k2_C_3C_O}
\end{center}
\end{figure}

In addition to its dependence on metallicity, the sensitivity to $\zeta_{\rm He}$/$\zeta_{\rm H_2}$ depends strongly on the C/O elemental abundance ratio. In Fig.~\ref{k1k2_C_3C_O}, we show the steady-state abundance of atomic carbon as a function of $\zeta_{\rm He}$/$\zeta_{\rm H_2}$ computed for the three sets of elemental abundances, each with three different values of C/O  -- 0.2, 0.4 (the standard one), and 0.6 -- obtained by maintaining a constant carbon abundance while varying the oxygen. The lower the C/O ratio is, the more the inflection point is shifted towards higher values of $\zeta_{\rm He}$/$\zeta_{\rm H_2}$. As an example, the inflection point for the high-metal case occurs around 0.2 for C/O=0.6 whereas it rises to 1.5 for C/O=0.2. In both cases, the inflection point is too far from the standard value of $\zeta_{\rm He}$/$\zeta_{\rm H_2}$ to obtain a bimodal distribution. For the intermediate-metal case, the C/O ratio of 0.6 results in shifting the inflection point to the standard value of $\zeta_{\rm He}$/$\zeta_{\rm H_2}$, so that the analogous panel to the one in Fig.~\ref{dist_C_3CI} would show two equal areas for the two peaks. The dispersion of the carbon abundance caused by the variation of C/O is larger for values of $\zeta_{\rm He}$/$\zeta_{\rm H_2}$ smaller than the inflection point while the abundances eventually form a single plateau after the jump.

\begin{figure}
\begin{center}
\includegraphics[width=1\linewidth]{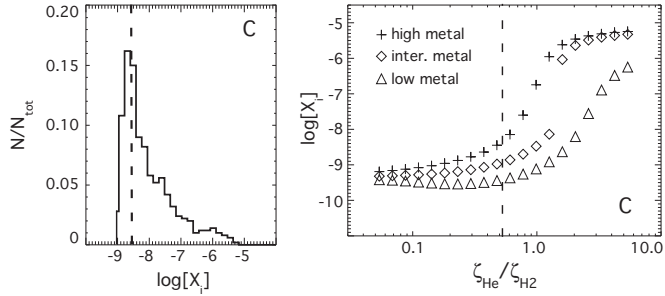}
\caption{The left panel shows the distribution of the steady-state abundance of atomic carbon computed with a random variation of $\zeta_{\rm He}$ and $\zeta_{\rm H_2}$ for the high-metal elemental abundances but with a lower helium elemental abundance of 0.18 compared with H$_2$. The right panel represents the C steady-state abundance as a function of $\zeta_{\rm He}$/$\zeta_{\rm H_2}$ for the three elemental abundances with the low He abundance. The vertical dashed line indicates the C abundances computed with the standard $\zeta_{\rm He}^0$ and $\zeta_{\rm H_2}^0$ on the left plot and the standard value of $\zeta_{\rm He}^0$/$\zeta_{\rm H_2}^0$ on the right plot.}
\label{dist_C_lowHe}
\end{center}
\end{figure}

As discussed in Wakelam \& Herbst (in preparation), the helium abundance used in the classic low- and 
high-metal elemental abundances (see Section~\ref{model}) is quite high (0.28 compared with H$_2$) compared with the more modern value of 0.18 measured in the interstellar medium \citep[see, for example,][]{1991ApJ...374..580B}.  Naturally, the sensitivity will change if a lower value of He is used. Fig.~\ref{dist_C_lowHe} shows the distribution of the steady-state C abundance computed with a random variation of $\zeta_{\rm He}$ and $\zeta_{\rm H_2}$ and the high-metal case with the lower helium abundance. In this case, we do not obtain a clear bimodal distribution but merely an asymmetrical profile, because the inflection point is shifted towards higher values of $\zeta_{\rm He}$/$\zeta_{\rm H_2}$, as shown in  Fig.~\ref{dist_C_lowHe},  and the random method barely reaches the sensitive portion of the  curve. 

\subsection{Chemical database}

\begin{figure}
\begin{center}
\includegraphics[width=1\linewidth]{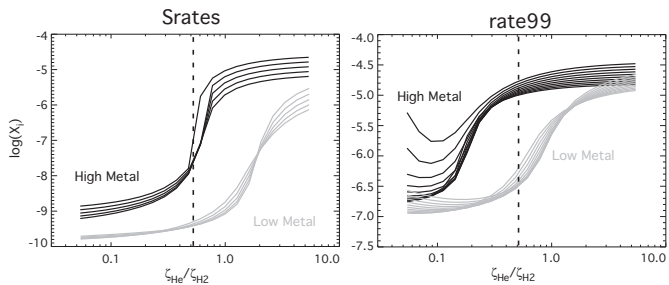}
\caption{Abundance of C as a function of $\zeta_{\rm He}$/$\zeta_{\rm H_2}$ computed with the high- (black lines) and low- (grey lines) metal abundances. The left and right panels show the results using the {\it Srates} (see text) and rate99 networks. The vertical dashed line indicates the standard value  $\zeta_{\rm He}^0$/$\zeta_{\rm H_2}^0$.}
\label{k1k2_echant_O2_Autre}
\end{center}
\end{figure}

To test the influence of the chemical network utilized, we constructed the sensitivity curves at steady-state with two other databases -- {\it Srates} and rate99-- using both low-metal and high-metal abundances. The rate99 network is the UMIST list of reactions \citep{2000A&AS..146..157L} prior to its current updating (see http://www.udfa.net) whereas {\it Srates} is the list of reactions used by \citet{2004A&A...422..159W} to study sulphur chemistry in hot cores. This latter network contains only five elements (H, He, C, O and S) but, as discussed in \citet{2004A&A...422..159W}, the steady-state abundances computed with it are very close to the ones computed with the larger osu.2003 network at low temperature. The ionization rates for H$_2$ and He are the same in all three databases.

Fig.~\ref{k1k2_echant_O2_Autre} is a sensitivity plot of the C abundance  for five different values of $\zeta_{\rm He}$ between 2$\zeta_{\rm He}$ and $\zeta_{\rm He}$/2.
The sensitivity obtained with {\it Srates} is similar to that for osu.2003, which indicates that the absence of the other elements (Si, Fe, Na etc.) does not change the sensitivity drastically. On the contrary, the results with rate99  show a much different relation between $\zeta_{\rm He}$/$\zeta_{\rm H_2}$ and the atomic carbon abundance, especially for the high-metal case, where the rate99 curves show (a) a significantly lower value for the point of inflection of the sensitivity, (b) a wide dispersion due to the value of $\zeta_{\rm He}$ at low values of $\zeta_{\rm He}$/$\zeta_{\rm H_2}$,  and (c) a much smaller  jump in the abundance.  We have shown in \citet{2006A&A...451..551W} that rate99 and osu.2003 often have differences in rate coefficients larger than the estimated random errors for unstudied reactions of a factor of 2, whereas most of the rate coefficients in {\it Srates} are taken from the osu database. Since the three networks have the same $\zeta_{\rm He}$ and $\zeta_{\rm H_2}$, the shape of the sensitivity can depend only on the values of other rate coefficients. 

\section{Bistability}\label{bistability}

\begin{figure}
\begin{center}
\includegraphics[width=1\linewidth]{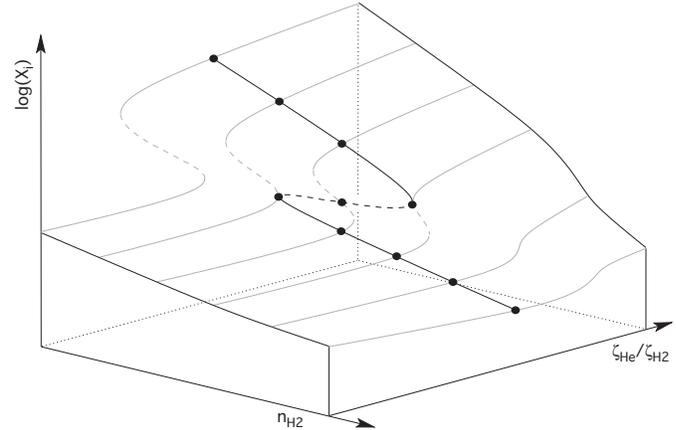}
\caption{Illustrative scheme of  hysteresis surface demonstrating bistability with two ``orthogonal'' control parameters, here $\rm n_{H_2}$ and $\zeta_{\rm He}$/$\zeta_{\rm H_2}$). The position of unstable solutions is indicated by dashed lines .  Orthogonal cuts crossing the folding region produce hysteresis curves like the ones shown in Fig.~\ref{hyst_nH2_C} and Fig.~\ref{hyst_nH2_k1k2_E}.}
\label{Bistable2ParametersDots}
\end{center}
\end{figure}

\begin{figure}
\begin{center}
\includegraphics[width=1\linewidth]{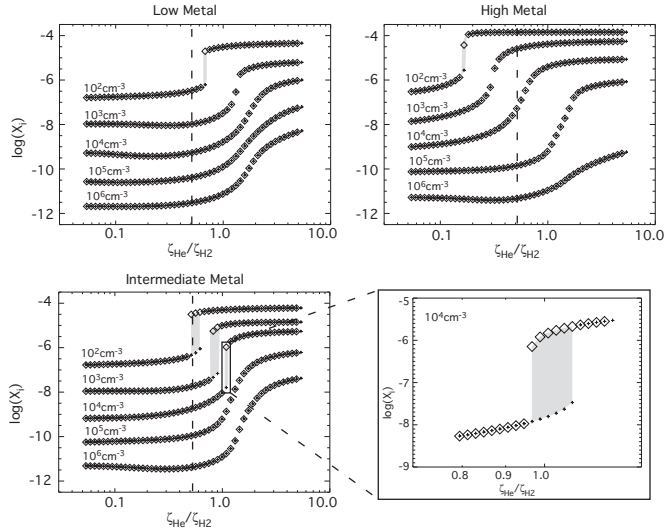}
\caption{Hysteresis curves of the steady-state abundance of atomic carbon as a function of $\zeta_{\rm He}$/$\zeta_{\rm H_2}$ for three different elemental abundances (low-, high-, and intermediate-metal), and for five H$_2$ densities between $10^2$ and $10^6$~cm$^{-3}$. The vertical dashed line indicates the standard value of $\zeta_{\rm He}^0$/$\zeta_{\rm H_2}^0$.  Grey zones indicate regions of bistability.}
\label{hyst_nH2_C}
\end{center}
\end{figure}

\begin{figure}
\begin{center}
\includegraphics[width=1\linewidth]{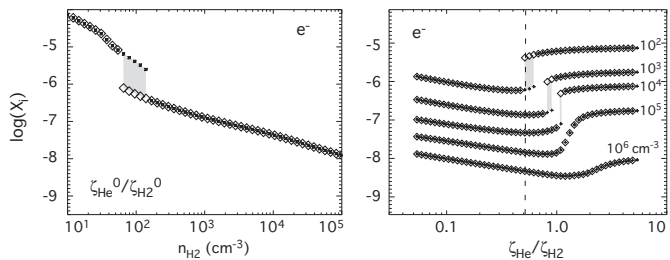}
\caption{Hysteresis curves of the electron steady-state abundance as a function of $n_{\rm H_2}$ (left panel) and $\zeta_{\rm He}$/$\zeta_{\rm H_2}$ (right panel). Intermediate-metal elemental abundances are used for both plots. Five H$_2$ densities between $10^2$ and $10^6$~cm$^{-3}$ are used for the right panel. The vertical dashed line indicates the standard value of $\zeta_{\rm He}^0$/$\zeta_{\rm H_2}^0$. Grey zones indicate regions of bistability. }
\label{hyst_nH2_k1k2_E}
\end{center}
\end{figure}

In section 4, we saw that the variation of $\zeta_{\rm He}$/$\zeta_{\rm H_2}$ leads to a sharp change in the abundances of some species for the intermediate-metal case (see Fig.~\ref{k1k2_3CI}). This jump is indeed a manifestation of the same bistability discovered by \citet{1993ApJ...416L..87L}, as can be determined by comparing the region of the bistability in parameter space. The parameter usually used to study this bistability is the ratio between the cosmic-ray ionization rate $\zeta$ and the density $n_{\rm H}$ but other model parameters such as the temperature and rate coefficients have been used \citep{1995A&A...297..251L,2000RSPTA.358.2549P}. What we have found here is that $\zeta_{\rm He}$/$\zeta_{\rm H_2}$ is also a control parameter which explore another dimension of the bistability. If we imagine $\zeta_{\rm He}$/$\zeta_{\rm H_2}$ to be an orthogonal axis to $\zeta$/$n_{\rm H}$, the bistability of \citet{1993ApJ...416L..87L} is the same as ours when 
$\zeta_{\rm He}$/$\zeta_{\rm H_2}$ is at its standard value.   In Fig.~\ref{Bistable2ParametersDots}, we show an illustrative scheme of this bistability as controlled by the two parameters $\rm n_{H_2}$  and $\zeta_{\rm He}$/$\zeta_{\rm H_2}$.
To make the bistability appear, we constructed hysteresis curves as a function of $\zeta_{\rm He}$/$\zeta_{\rm H_2}$ for different H$_2$ densities. These curves, shown in Fig.~\ref{hyst_nH2_C} for atomic carbon, were produced in two steps. Fist, we ran the model to compute the steady-state abundance using one value of $\zeta_{\rm He}$/$\zeta_{\rm H_2}$, then we used this solution as the initial condition and increased $\zeta_{\rm He}$/$\zeta_{\rm H_2}$ to compute the next point, shown as the crosses in Fig.~\ref{hyst_nH2_C} and \ref{hyst_nH2_k1k2_E}. In addition, the same principle but with decreasing $\zeta_{\rm He}$/$\zeta_{\rm H_2}$ was utilized, with results shown as  diamonds..  In order to vary $\zeta_{\rm He}$/$\zeta_{\rm H_2}$, we fixed the helium ionization rate and changed that of hydrogen.  When two solutions exist for a range of $\zeta_{\rm He}$/$\zeta_{\rm H_2}$ values,  a region of bistability exists.

To explore the domains of bistability using the new control parameter, we constructed  hysteresis curves using the three different elemental abundances and five densities between $10^2$ and $10^6$~cm$^{-3}$. All the results are shown in Fig.~\ref{hyst_nH2_C} for atomic carbon. Bistability exists for all the elemental abundances but the domain of $\zeta_{\rm He}$/$\zeta_{\rm H_2}$ where the two solutions coexist are small.  To show the bistability more clearly,  a magnification of this region is done for n(H$_2$)=10$^4$~cm$^{-3}$ and the intermediate-metal case. In both the low- and high-metal cases,  bistability exists only at very low density ($10^2$~cm$^{-3}$) whereas in the intermediate-metal case, the bistability exists between $10^2$ and $10^4$~cm$^{-3}$. The increase in density shifts and smoothes the inflection point, making the bistability eventually disappear.  Note that this is the first time that the bistability has been found with the high-metal case \citep{1998A&A...334.1047L,2006ApJ...645..314B}, because it does not exist for the standard value of  $\zeta_{\rm He}$/$\zeta_{\rm H_2}$.  We have confirmed that, for the high-metal case, using the standard value of $\zeta_{\rm He}$/$\zeta_{\rm H_2}$ leads to no bistability with $\rm n_{H_2}$ as low as 10~cm$^{-3}$. Using $\zeta_{\rm He}$/$\zeta_{\rm H_2} = 0.16$ however, bistability exists in a very small density range of 70-90~cm$^{-3}$.

A more focused look at the two control parameters is shown in  Fig.~\ref{hyst_nH2_k1k2_E}, where we plot the hysteresis curves of the electron abundance as a function of each parameter for the case of intermediate-metal elemental abundances.  The value of $\zeta$ is set at the standard value. For the variation of $\zeta_{\rm He}$/$\zeta_{\rm H_2}$ (right panel), we used five H$_2$ densities between $10^2$ and $10^6$~cm$^{-3}$, as in the previous figure, while for the hysteresis curve as a function of $\rm n_{H_2}$ (left panel) the standard value of $\zeta_{\rm He}$/$\zeta_{\rm H_2}$ (0.54) was used and the bistability found to exist in the density range 60-150~cm$^{-3}$. With the other control parameter and for  $\rm n_{H_2} = 10^2$~cm$^{-3}$, the bistability exists for $\zeta_{\rm He}$/$\zeta_{\rm H_2}$ between 0.5 and 0.7 and the domains of bistability using both parameters overlap. For higher densities, the bistable region using $\zeta_{\rm He}$/$\zeta_{\rm H_2}$ shifts towards higher values and so does not exist anymore in the hysteresis curve versus $\rm n_{H_2}$.  These hysteresis curves can be combined into a plot of the type shown in Fig.~\ref{Bistable2ParametersDots}.

\section{Chemical reactions involved in the sensitivity}


The sensitivity of the abundances to the ratio $\zeta_{\rm He}$/$\zeta_{\rm H_2}$ is related to the non-linearity of the system. A detailed mathematical analysis of this phenomenon is in progress (Massacrier et al., in preparation). It is possible, however, to understand the difference,  between the low and high $\zeta_{\rm He}$/$\zeta_{\rm H_2}$ phases qualitatively from a chemical point of view.  For this purpose, we have considered the main reactions of production and destruction of 
O, C, CO, S and SO, which are the main reservoirs of carbon, oxygen and sulphur for the intermediate- and low-metal elemental compositions in which sulfur is relatively abundant, and/or are the species  most influenced by $\zeta_{\rm He}$ and $\zeta_{\rm H_2}$.
Let us first consider low values of $\zeta_{\rm He}$/$\zeta_{\rm H_2}$.  The ionization of H$_2$ leads to the formation of OH through a well-known series of reactions \citep[see. e.g. ][]{2006ApJ...645..314B}. The radical OH then reacts with atomic oxygen to form O$_2$. The most efficient way of destroying O$_2$ is to form SO by reaction with S. Since the elemental abundance of atomic sulphur is much less than that of oxygen, large amounts of O$_2$ remain in the gas phase. Atomic carbon reacts with O$_2$ to form CO where it is stored since CO is not efficiently destroyed.  For high $\zeta_{\rm He}$/$\zeta_{\rm H_2}$, on the other hand, the destruction of CO by He$^+$  produces C$^+$ and O more efficiently. Carbon ion then exchanges its charge with S so large abundances of C remain in the gas phase to deplete molecular oxygen. 

Since the bistability we see is an extreme manifestation of the sensitivity to $\zeta_{\rm He}$/$\zeta_{\rm H_2}$, the chemical processes involved in bistability may be related to those discussed above, although our control parameter is different from those previously studied. \citet{1992MNRAS.258P..45P} and \citet{1993ApJ...416L..87L,1995A&A...297..251L} identified the abundance ratio H$^+$/H$_3^+$ as being important for bistability, with one ion identified with the HIP and the other with the LIP solution. The abundance ratio H$^+$/H$_3^+$ depends on the abundance of electrons and as a consequence on the efficiency of the dissociative recombination of H$_3^+$ with electrons, the rate of which was quite uncertain at the time.   In our analysis however, the H$_{3}^{+}$ ion shows little if any sensitivity to the control parameter, although the electron abundance does show a jump between the low and high $ \zeta_{\rm He}$/$\zeta_{\rm H_2}$ phases, which can then be thought of as analogous in some sense  to  the LIP and HIP solutions.  Recently, \citet{2006ApJ...645..314B} made another detailed analysis of the chemical differences between the so-called HIP and LIP solutions, arguing that bistability is due to a loop including H$_3^+$, O$_2$ and S$^+$.  Although this explanation is related in some aspects to that given for the sensitivity reported here, one must remember that our control parameter raises some new points, given the relation of the ionization rate of He to the formation of the important ion C$^{+}$, which is a precursor for many species, and can charge exchange to form abundant C.  Moreover, S$^{+}$, which is prominent in the explanation of \citet{2006ApJ...645..314B}, is not the dominant charge carrier in our calculations.  Note that the osu.2003 database includes the recombination of the main ions with negative grains and this does not cancel the bistability, although such a cancellation was
suggested by \citet{2006ApJ...645..314B}.
It may well be that the H$^+$/H$_3^+$ abundance ratio,  the H$_3^+$-O$_2$-S$^+$ cycle, and the $\zeta_{\rm He}$/$\zeta_{\rm H_2}$ ratio comprise pieces of a larger and more complex jigsaw puzzle. Since bistability also appears to be a function of the reaction network utilized, secondary reactions may play a more important role than heretofore realized.
This possibility has to be clarified through the building of a simplified reaction network that still shows the main characteristics of bistability.

\section{Observational relevance}

The sensitivity to $\zeta_{\rm He}$/$\zeta_{\rm H_2}$ depends highly on the model parameters and, for this reason, it is not obvious if such effect can be detected in the interstellar medium.  In an ideal situation where the cloud characteristics  (at least the elemental abundances and the age) are well known and correspond to a distinct chemical regime, one could use the observations to constrain the ratio $\zeta_{\rm He}$/$\zeta_{\rm H_2}$ assuming however that no local variation of this ratio is expected. In reality, the cloud characteristics are in general far from being well constrained, and an analysis of the uncertainties in the ionization rates of H$_2$ and He is strongly required to understand the relevance of this sensitivity for the interstellar medium. Note that $\zeta_{\rm He}$/$\zeta_{\rm H_2}$ is not sensitive to the total intensity of the cosmic-ray flux. 

If the intermediate-metal elemental abundances reflect the  cloud characteristics and if the ionization rate of He and H$_2$ by cosmic-rays can vary by at least a factor of 2 within a cloud, the prediction of bistability over a wide range of densities suggests that  strong chemical heterogeneities in dense clouds can be caused by this effect.  In the absence of bistability an extreme sensitivity is predicted at most densities for low-metal and high-metal abundances and may cause some observed heterogeneities in abundance. However, we have to consider the age of the cloud. The bifurcation occurs just after $10^5$~yr, starting from our standard initial conditions, and is a maximum at steady-state ($\geqslant 10^7$~yr). As a consequence, the chemical heterogeneity can be caused by this sensitivity only in evolved clouds ($\geqslant 10^5 - 10^6$~yr). 

The large gas-phase abundances of H$_2$O and O$_2$ molecules predicted by dense cloud chemical models is a long standing problem. Using high values of $\zeta_{\rm He}$/$\zeta_{\rm H_2}$, we obtain much smaller abundances for these species, closer to the upper limits in dense clouds \citep{2002ApJ...581L.105B,2003A&A...402L..77P}. Large uncertainties in the $\zeta_{\rm He}$/$\zeta_{\rm H_2}$ ratio could then solve the problem related to the modeling of water and molecular oxygen. 
One has, however, to consider the overall picture, and the better agreement with H$_2$O and O$_2$ should not worsen the agreement for the other species. A comparison between observations and modeling should be done in a systematic way, as done by \citet{2006A&A...451..551W}. This was not the purpose of the present study, but we can already note, as an example, that the observed abundances in L134N (North peak) of OH and C$_4$H \citep[$7.5\times 10^{-8}$ and $10^{-9}$ respectively,][]{1992IAUS..150..171O} are reproduced by the model when using the standard value of $\zeta_{\rm He}$/$\zeta_{\rm H_2}$ but are underestimated when using the higher ratios $\zeta_{\rm He}$/$\zeta_{\rm H_2}$ that reproduce H$_2$O and O$_2$ abundances. 

Finally, the very low H$_2$ density (100~cm$^{-3}$), at which the bistability as a function of $\zeta_{\rm He}$/$\zeta_{\rm H_2}$ is restricted to with the low- and high-metal elemental abundances, is not relevant for  clouds with the assumed visual extinction ($A_v$=10) since it would require a size of 26~pc, which is much larger than typical cloud sizes \citep{1981ApJS...45..121S}.

\section{Conclusions}

Using a Monte Carlo uncertainty method to compute the theoretical error of molecular abundances in chemical models, we found a strong sensitivity of the steady-state abundances to the ratio between the ionization rate coefficient of helium $\zeta_{\rm He}$ and that of molecular hydrogen $\zeta_{\rm H_2}$ by cosmic rays. This sensitivity to the ratio has the consequence of changing the abundance of some species by several orders of magnitude for small variations of $\zeta_{\rm He}$/$\zeta_{\rm H_2}$. Lower values of $\zeta_{\rm He}$/$\zeta_{\rm H_2}$ are characterized by large abundances of smaller molecules such as O$_2$, H$_2$O, SO, etc. whereas high values of $\zeta_{\rm He}$/$\zeta_{\rm H_2}$ lead to large abundances of atoms. If the sensitivity is strong enough, it can result in the phenomenon of bistability, which we located at low densities for two sets of elemental abundances - the low- and high-metal cases, and over a wide range of densities for the case of the so-called intermediate-metal abundances, in which sulfur plays a particularly prominent role.  The bistability has been determined to be the same as studied earlier by a variety of authors with other control parameters; it is perhaps simplest to consider our control parameter --  $\zeta_{\rm He}$/$\zeta_{\rm H_2}$-- as orthogonal to one of the others - the ratio of the overall ionization rate parameter $\zeta$ to the density $n$ --  so that the bistability can be imagined to exist in a region in three-dimensional space rather than in the customary two-dimensional hysteresis picture.  

It is interesting to speculate that the sensitivity detected here via our uncertainty analysis is a necessary but not sufficient condition for bistability because only the most extreme sensitivity, in which the bimodal distributions detected have no solutions in between them, leads to bistability. Unfortunately, it is not guaranteed that our type of uncertainty analysis can always locate a bistable region since finding an extreme sensitivity may require large uncertainties in some parameters, such as the gas density, to take us from our standard parametric values to the bistable region.  

Several questions arise when considering the dramatic effects of small $\zeta_{\rm He}$/$\zeta_{\rm H_2}$ variations on the gas-phase chemistry. First, even if one assumes a given flux and energy distribution of the cosmic rays, the determination of $\zeta_{\rm He}$/$\zeta_{\rm H_2}$ suffers some significant uncertainties, with the role of secondary electrons particularly important.  Besides these  difficulties, one might also have to consider variations of the ratio $\zeta_{\rm He}$/$\zeta_{\rm H_2}$ in space, such as between one cloud and another, or between the center and the edges of a same cloud,  and in time. Changes in the ratio imply important variations in the spectral distribution of the cosmic rays \citep{1973ApJ...186..859G}. Such variations might be produced by the attenuation of the cosmic rays inside a cloud or  the proximity of supernovae in star forming clouds. Whether such variations are expected or not and how they could affect the chemistry should be considered in the future. Finally, a knowledge of the helium elemental abundance appears to be crucial since the sensitivity region depends on this parameter. 

\begin{acknowledgements}

V. W. and E. H. thank the National Science Foundation for its partial support of this work. G.M. acknowledges partial financial support from the CNRS/INSU programs PCMI and PNPS.

\end{acknowledgements}

\bibliographystyle{aa}

\end{document}